\documentclass[12pt]{iopart}
\usepackage[bookmarks=false]{hyperref}

\usepackage{iopams}
\usepackage{subfig}  
\usepackage{graphicx}
\usepackage{cite}
\begin{document}

\title{Optical qubit generation via atomic postselection in a Ramsey interferometer}

\author{M Orszag $^{1,2}$ and F Oyarce$^1$}

\address{$^1$ Instituto de F\'{i}sica, Pontificia Universidad Cat\'{o}lica de Chile, Casilla 306, Santiago, Chile}
\address{$^2$ Universidad Mayor, Avda. Alonso de C\'{o}rdova 5495, Las Condes, Santiago, Chile}
\ead{miguel.orszag@umayor.cl, fioyarce@uc.cl}
\vspace{10pt}
\begin{indented}
\item[]July 2018
\end{indented}

\begin{abstract}
We propose a realizable experimental scheme to prepare a superposition of the vacuum and one-photon states using a typical cavity QED-setup. This is different from previous schemes, where the superposition state of the field is generated by resonant atom-field interaction and the cavity is initially empty. Here, we consider only dispersive atom-field interaction and the initial state of the cavity field is coherent. Then, we determine the parameters to prepare the desired state via atomic postselection. We also include the effect of cavity losses and detection imperfections in our analysis, against which this preparation of the optical qubit in a real Fabry-P\'{e}rot superconducting cavity is robust. Additionally, we show that this scheme can be used for the preparation of other photon number Fock state superpositions. In summary, our task is achieved with a high fidelity and a postselection probability within experimental reach
\end{abstract}

%
%
%
%
%

\section{Introduction}
Generation and engineering of nonclassical states of light is central to quantum optics and quantum information. Over the years, various schemes for the preparation of Fock states \cite{krause, QND90, QND91, QND92, leonski, leonski2, domokos, varcoe, dotsenko} and their arbitrary finite superpositions \cite{voguel, moussa, PPB, Barnett, dakna, paris, dariano, serra, qsd2001, qsd2002, nonlinearscissors} have been developed. Such states have been shown to be generated by nonlinear media or by conditional measurements. For example, the method proposed by Brune {\it et al.} \cite{QND90, QND92} can generate Fock states of a cavity field. This method it is based on a quantum nondemolition measurement (QND) of the photon number of a field stored in a high-Q cavity, where the information acquired by detecting a sequence of atoms modifies the field step by step, until it eventually collapses into a Fock state. This has been done experimentally by Guerlin {\it et al} \cite{guerlin}. Although the collapse of the field into a Fock state it is not predictable due to the randomness of the measurement, it is posible to prepare on demand photon number states (Fock states) using a quantum feedback scheme in the context of a QND measurement of the photon number of a cavity field \cite{dotsenko, sayrin}. Other interesting schemes proposed by Leo\'{n}ski \cite{leonski, leonski2} relies on the nonlinearity of the time evolution of the cavity field in a Kerr medium. In these methods, by adjusting the parameters of the Hamiltonian it is possible to generate a Fock state via unitary time evolution of a given initial cavity field.

In the case of the preparation of a finite superposition of the number state, most of the proposed schemes are based on a conditional measurement at the outports of beam splitters. For example, the method proposed by Dakna {\it et al.} \cite{dakna} generates an arbitrary (finite) superposition of Fock states by performing alternaly coherent displacement and single-photon adding in a well-defined succesion via conditional measurements on beam splitters. However, one of the simplest methods is the optical truncation of coherent light, also referred to as {\it quantum scissors device} (QSD), proposed by Pegg, Phillips and Barnett (PPB) \cite{PPB, Barnett}. Later, Resch {\it et al.} proposed and experimentally demonstrated a QSD-like state preparation technique based on conditional coherence \cite{resch}. Also, due to its simplicity, the basic idea of the QSD has been modified and generalized \cite{paris}. It was shown that an optical qubit can be generated experimentally with high fidelity using the PPB scheme with commercially available detectors and single-photon sources \cite{qsd2001, qsd2002}. Moreover, optical qubit generation by nonlinear quantum scissors has been proposed \cite{nonlinearscissors}.
Aditionally, various schemes has been presented in the realm of cavity QED for the generation of superpositions states of the radiation field by a conditional measurement of the atoms. Voguel {\it et al.} have basically employed resonant atom-field interaction to build an arbitrary field in an initial empty cavity \cite{voguel}. Another proposal \cite{moussa}, based on \cite{voguel}, has been presented considering both resonant and dispersive atom-field interaction for preparing a reciprocal-binomial state of the radiation field from an initial empty cavity. Also, a different method considers an initial coherent state and two Ramsey zones for the generation of a general cavity field state by the conditional measurement of the atoms interacting resonantly with the field \cite{serra}.

In this paper, our main interest is to propose and study a typical cavity QED-setup used for Ramsey interferometry to produce a superposition of the vacuum and one-photon states, which is the simplest optical qubit state. The main idea is based on a initial preparation of atoms that enter in a Ramsey interferometer and whose final states are postselected, adjusting the various parameters in such a way that we can generate an optical qubit. The principal difference with the previous schemes presented in cavity QED is that we consider only dispersive atom-field interaction for the generation of this kind of superposition state. Hence, there is no energy exchange between the field and the atoms. Moreover, this scheme can be employed for create other photon number Fock states.

This article is organized as follows. In section 2, we present our model and derive the general expression of the cavity field after atomic postselection. In section 3, we focus our work on the preparation of an optical qubit (vacuum and one-photon state superposition) from an initial coherent state of the cavity field. Here, we determine the parameters which optimize the fidelity between the final and the target state, and we calculate the postselection probability. In section 4, we describe a real experiment where our scheme can be done using a microwave QED system with circular Rydberg atoms. We discuss the feasibility of our scheme under real experimental conditions and calculate the effect of cavity losses and detection imperfections in the generation of the optical qubit. Aditionally, we show that our scheme is not limited for the preparation of superpositions of vacuum and one-photon states and we present some examples of other photon number Fock states superpositions. Finally, we present the concluding remarks of our research.
\section{The model}
\begin{figure}[h]
\centering \includegraphics[width=\linewidth]{setup.pdf}
\centering \includegraphics[scale=0.2]{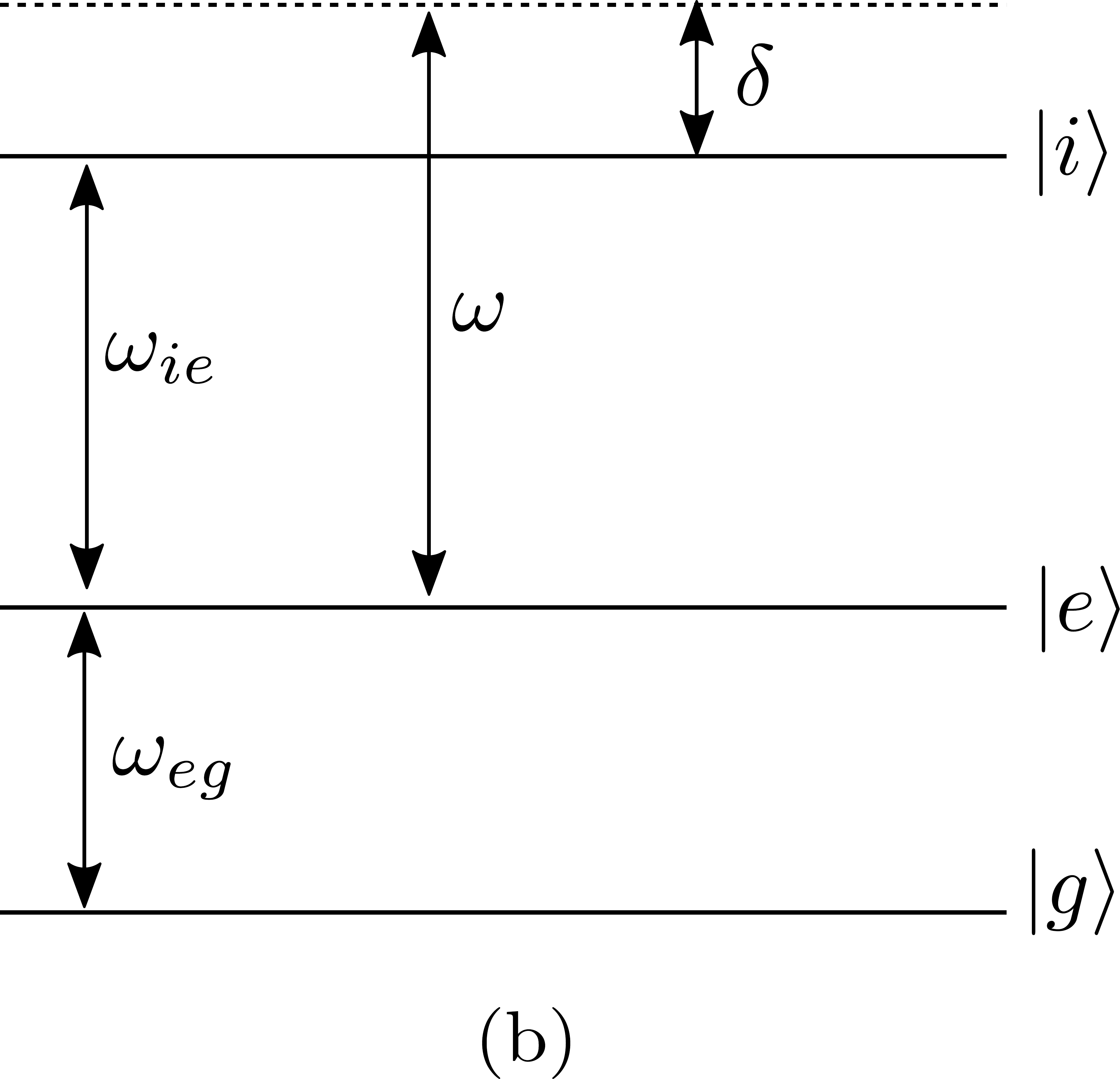}
\caption{(a) Cavity QED-setup used for Ramsey interferometry. The field is initially prepared in the high-Q cavity $C$. The atoms are prepared and velocity-selected in the box $O$. Then, each atom interacts with three cavities: $R_1$, $C$ and $R_2$. In each of the zones $R_1$ and $R_2$, the atom interacts with a classical microwave field. This interaction makes it possible to manipulate the atomic state before and after the interaction with $C$.  Finally, after passing zone $R_2$, the atom is detected in the state $|e\rangle$ or $|g\rangle$ by the field ionization counter $D$. (b) Three-level atomic system for the experiment in the dispersive atom-field coupling. Here, $\omega$ is the frequency of the field in cavity $C$ which has a large detuning $\delta$ from the atomic transition frequency $\omega_{ie}$.}
\label{figure0}
\end{figure}
This work is based on the detection of the dispersive phase shift caused by the cavity field on the wave function of nonresonant atoms crossing the cavity \cite{QND90}. The cavity QED-setup to measure this shift by Ramsey interferometry is shown in figure \ref{figure0}a. On the one hand, the initial state of the field (with a well-known photon distribution) is prepared in the cavity $C$ between two classical microwave zones $R_{1}$ and $R_{2}$. On the other hand, $N$ succesive three-level atoms are initialized in the $|e\rangle$ level, with the energy diagram shown in figure \ref{figure0}b. These atoms are injected into the setup at a very low rate such that there is only one atom in the cavity at a given time. Thus, the initial state of the combined atoms-cavity system is given by
\begin{equation}
\rho_{ca} = \rho_{c}\otimes\rho_{a} = \sum_{nn'}\rho_{nn'}|n\rangle\langle n'|\otimes |e_{1}\rangle\langle e_{1}|\otimes ...\otimes|e_{N}\rangle\langle e_{N}|,
\end{equation}
where the cavity field is spanned in Fock basis $|n\rangle$, and $|e_{k}\rangle$ represents the $k$th atom. In the following we describe the main features of the cavities $R_1$, $C$ and $R_2$ and the evolution of the atom-field system while the atom crosses these cavities.

Generally, in cavity QED, the atom-field interaction is described by the Jaynes-Cummings model
\begin{equation}
H^{(k)} = \frac{\hbar}{2}\omega_{ie}\sigma_{z}^{(k)}+\hbar\omega a^{\dag}a+\hbar g(a\sigma_{+}^{(k)}+a^{\dag}\sigma_{-}^{(k)}),
\end{equation}
where $a$ ($a^{\dag}$) is the cavity photon annihilation (creation) operator, whereas for the $k$th atom the operators are $\sigma_{-}^{(k)}=|e_{k}\rangle\langle i_{k}|$, $\sigma_{+}^{(k)}=|i_{k}\rangle\langle e_{k}|$, $\sigma_{z}^{(k)}=|i_{k}\rangle\langle i_{k}|-|e_{k}\rangle\langle e_{k}|$ and $g$ corresponds to the atom-field coupling constant, taken to be equal for all the atoms. From the atomic operators it can be seen that only levels $|i_{k}\rangle$ and $|e_{k}\rangle$ are affected by the atom-field interaction, whereas level $|g_{k}\rangle$ is not involved in the dynamics. Particularly, in our scheme, we are interested in considering a nonresonant interaction by taking a large frequency detuning $\delta = \omega_{ie}-\omega\gg$ $g\sqrt{n}$ between the cavity field frequency $\omega$ and the atomic transition frequency $\omega_{ie}$. Therefore, the effective interaction becomes a dispersive coupling \cite{scully}
\begin{equation}
V ^{(k)}= \frac{\hbar g^{2}}{\delta}a^{\dag}a|e_{k}\rangle\langle e_{k}|.
\end{equation}

After an interaction time $\tau_{k} = L/v_{k}$, the evolution operator reads as
\begin{equation}
U_{I}^{(k)}= \exp(-\rmi V^{(k)}\tau_{k}/\hbar)= \exp(-\rmi\varphi_{k} a^{\dag}a|e_{k}\rangle\langle e_{k}|).
\end{equation}

In the above, $L$ is the length of the cavity $C$, $v_{k}$ is the velocity of the $k$th atom passing through the cavity and $\varphi_{k}=g^{2}\tau_{k}/\delta$ is the one photon phase shift, which caracterizes the coupling strength between the $k$th atom and the cavity field. This interaction causes a dispersive phase shift to the $|e_{k}\rangle$ level which is proportional to the photon number.

On the other hand, in each of the $R_{1}$ and $R_{2}$ zones, the $k$th atom interacts with a classical microwave field tuned at a frequency $\nu_{r}$, resonant with the atomic transition frequency $\omega_{eg}$. This interaction leads to a superposition of the $|e_{k}\rangle$ and $|g_{k}\rangle$ levels of the atomic state \cite{scully}. After an interaction time $\Delta\tau_{k}=\Delta L/v_{k}$, which satisfies $\Omega_{R}\Delta\tau_{k}=\pi/2$, the atom undergoes a $U_{\pi/2}^{(k)}$ transformation given by

\begin{equation}\label{u_pi}
U_{\pi/2}^{(k)}=\frac{1}{\sqrt{2}}(|e_{k}\rangle\langle e_{k}|+|g_{k}\rangle\langle g_{k}|+\rmi|e_{k}\rangle\langle g_{k}|+\rmi|g_{k}\rangle\langle e_{k}|),
\end{equation}
where $\Delta L$ is the length of the zones $R_{1}$ and $R_{2}$, $v_{k}$ is the velocity of the atom and $\Omega_{R}$ is the Rabi frequency.

The total evolution operator is given by
\begin{equation}
U = U^{(N)}...U^{(1)},
\end{equation}
being $U^{(k)}= U_{\pi/2}^{(k)}U_{I}^{(k)}U_{\pi/2}^{(k)}$ the evolution of the $k$th atom passing through the cavities ($R_{1}$, $C$ and $R_{2}$).

After the interaction of the $N$ atoms with the cavities, the state of the whole system evolves to
\begin{eqnarray}
\tilde{\rho}_{ca} &= U\rho_{ca}U^{\dag}
\\
&= \sum_{nn'}\rho_{nn'}|\psi_{n}^{(1)}\rangle\langle\psi_{n'}^{(1)}|\otimes...\otimes|\psi_{n}^{(N)}\rangle\langle\psi_{n'}^{(N)}|\otimes|n\rangle\langle n'|,
\end{eqnarray}
with $U^{(k)}|e_{k}\rangle=|\psi_{n}^{(k)}\rangle = \frac{1}{2}\left(\e^{-\rmi\varphi_{k} n}|e_{k}\rangle+\rmi\e^{-\rmi\varphi_{k} n}|g_{k}\rangle+\rmi|g_{k}\rangle-|e_{k}\rangle\right)$.\ For simplicity, we assume the same coupling $\varphi$ for all the atoms.

Subsequently, we perform the postselection of a symmetric state of the atomic levels on the $\left\{|m_{(1)},...,m_{(N)}\rangle\right\}$ basis, with $m = \left\{e,g\right\}$ 
\begin{eqnarray}
\nonumber|\phi_{post}\rangle &=|e_{1},...,e_{N_{e}},g_{N_{e}+1},...,g_{N}; S\rangle  
\\
&= \left(\frac{N_{e}!(N-N_{e})!}{N!}\right)^{1/2}\sum_{p}|m_{1},...,m_{N}\rangle.\label{postselection}
\end{eqnarray}

Here, $S$ stands for a symmetric state. Therefore, the sum is taken over all the possible combinations of $N_{e}$ atoms on the $|e\rangle$ level and $N-N_{e}$ on the $|g\rangle$ level.

Hence, the normalized state of the cavity after postselection is
\begin{eqnarray}
\nonumber\rho_{c}^{post}&=\frac{\langle\phi_{post}|\tilde{\rho}_{ca}|\phi_{post}\rangle}{\Tr_{c}\left[\langle\phi_{post}|\tilde{\rho}_{ca}|\phi_{post}\rangle\right]}
\\
&= \frac{C_{N}^{N_{e}}\sum_{nn'}\rho_{nn'}\rme^{-\frac{\rmi}{2}\varphi n N}\rme^{\frac{\rmi}{2}\varphi n' N}c_{n}^{N-N_{e}}c_{n'}^{N-N_{e}}d_{n}^{N_{e}}d_{n'}^{N_{e}}|n\rangle\langle n'|}{C_{N}^{N_{e}}\sum_{n}\rho_{nn}c_{n}^{2(N-N_{e})}d_{n}^{2N_{e}}},\label{poststate}
\end{eqnarray}
where $C_{N}^{N_{e}}=\frac{N!}{N_{e}!(N-N_{e})!}$ is the number of combinations of having $N_{e}$ atoms on the $|e\rangle$ level of a set of $N$ atoms, $c_{n} = \cos(\frac{\varphi n}{2})$ and $d_{n}=\sin(\frac{\varphi n}{2})$. The postselection probability is the denominator of equation (\ref{poststate}), defined as
\begin{equation}\label{postproba}
P_{post} = C_{N}^{N_{e}}\sum_{n}\rho_{nn}c_{n}^{2(N-N_{e})}d_{n}^{2N_{e}}.
\end{equation}

In what follows, we show that it is possible to generate an optical qubit with an appropiate atomic postselection. 

\section{Preparing an optical qubit in dispersive cavity-QED}
In this section, our task is to prepare a superposition of the vacuum and the one-photon states. First, we assume that the initial state of the field is a coherent state $|\alpha\rangle = \sum_{n}b_{n}|n\rangle$, where $b_{n}=\alpha^{n}\e^{-\left|\alpha\right|^{2}/2}/\sqrt{n!}$ and $\alpha$ being a real value. Thus, equation (\ref{poststate}) reduces straighforwardly for the case of an initial pure state for the cavity field, and for $N_{e}=0$ (because for $N_{e}\neq0$ the ket $|0\rangle$ is eliminated). The state of the field after the postselection of $N$ atoms in the $|g\rangle$ level is
\begin{equation}\label{finalstate}
|\psi_{f}\rangle = \frac{\sum_{n=0}^{\infty}b_{n}\e^{-\frac{\rmi n\varphi N}{2}}\cos^{N}\left(\frac{\varphi n}{2}\right)|n\rangle}{\left[\sum_{n=0}^{\infty}\left|b_{n}\right|^{2}\cos^{2N}\left(\frac{\varphi n}{2}\right)\right]^{1/2}},
\end{equation}
with postselection probablity
\begin{equation}\label{postprob}
P_{post}=\sum_{n=0}^{\infty}\left|b_{n}\right|^{2}\cos^{2N}\left(\frac{\varphi n}{2}\right). 
\end{equation}

As we can see from the numerator of equation (\ref{finalstate}), the parameters $\alpha$ , $\varphi$ and $N$ have to be adquate to ensure that only kets $|0\rangle$ and $|1\rangle$ survive.

We assume a target state of the form:
\begin{equation}\label{target}
|\psi_{t}\rangle= \frac{|0\rangle+\alpha \e^{-\frac{\rmi N\varphi}{2}}\cos^{N}\left(\frac{\varphi}{2}\right)|1\rangle}{\sqrt{1+\alpha^{2}\cos^{2N}\left(\frac{\varphi}{2}\right)}},
\end{equation}
which can be written simply as $|\psi_{t}\rangle
= \frac{|0\rangle+\beta|1\rangle}{\sqrt{1+|\beta|^2}}$, with $\beta=\alpha \e^{-\frac{\rmi\phi N}{2}}\cos^{N}\left(\frac{\varphi}{2}\right)$.

In order to estimate how far our final state $|\psi_{f}\rangle$ is from the target state $|\psi_{t}\rangle$, we define a fidelity $F=\left|\langle\psi_{t}|\psi_{f}\rangle\right|^{2}$ \cite{fidelity}. Using the equations (\ref{finalstate}) and (\ref{target}), the fidelity reads as
\begin{equation}\label{fidelity}
F = \frac{1+\alpha^{2}\cos^{2N}\left(\frac{\varphi}{2}\right)}{\sum_{n=0}^{\infty}\frac{\alpha^{2n}}{n!}\cos^{2N}\left(\frac{\varphi n}{2}\right)}.
\end{equation}

It can be easily seen from the above equation that a combination of $\alpha$, $\varphi$ and $N$ can lead to a specific optical qubit. Particularly, if we want to prepare an equiprobable superposition, e.g, $\left|\langle0|\psi_{t}\rangle\right|^{2}=\left|\langle1|\psi_{t}\rangle\right|^{2}=1/2$, we require
\begin{equation}\label{condition}
\alpha \cos^{N}\left(\frac{\varphi}{2}\right) = 1.
\end{equation} 

Solving this condition for the variable $\varphi$ gives
\begin{equation}\label{phi}
\varphi(N)=2\arccos\left(\frac{1}{\sqrt[N]{\alpha}}\right),
\end{equation}
where $\alpha$ is given. Replacing this condition into the expression of the fidelity in equation (\ref{fidelity}), we finally obtain a fidelity depending only on the number of atoms postselected in the $|g\rangle$ level given by
\begin{equation}\label{fidelity_N}
F(N) = \frac{2}{2+\sum_{n=2}^{\infty}\frac{\alpha^{2n}}{n!}\cos^{2N}\left(\frac{\varphi (N)n}{2}\right)}.
\end{equation} 
\begin{figure}[h]
\centering
\subfloat[]
{
    \includegraphics[scale=.5]{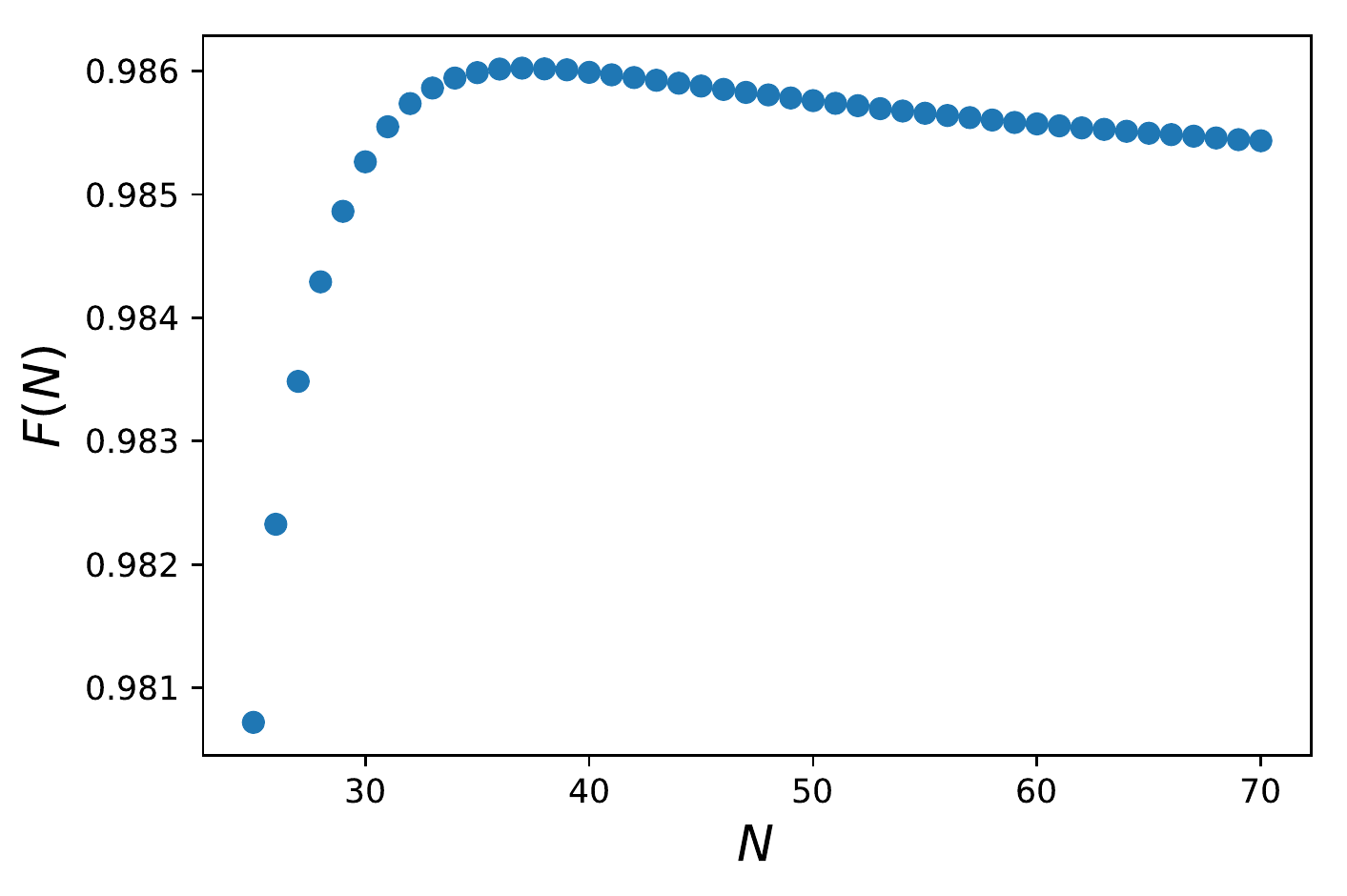}
    \label{fig:figure1a}
}
\subfloat[]
{
    \includegraphics[scale=.5]{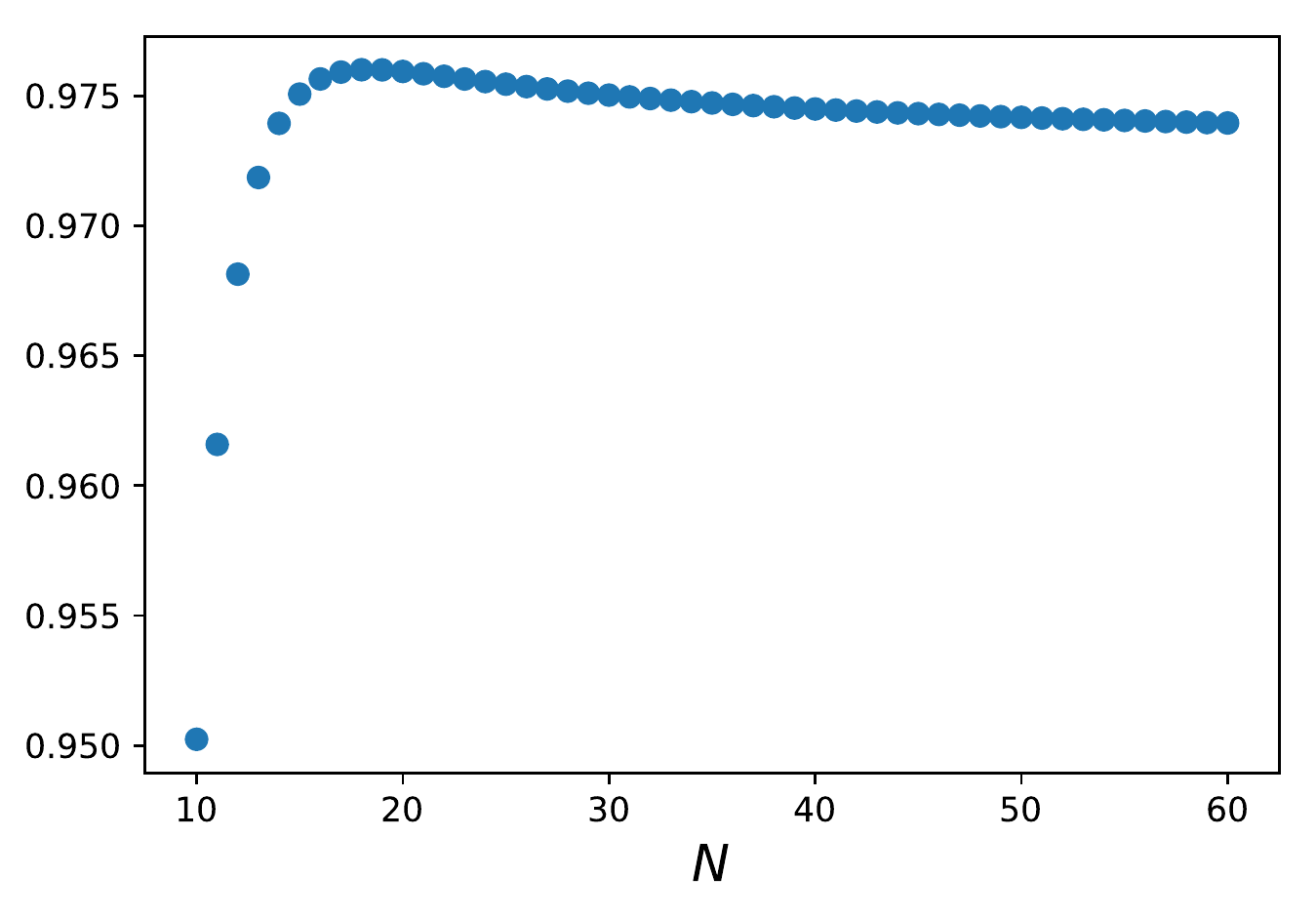}
    \label{fig:figure1b}
}
\caption{Fidelity given by equation (\ref{fidelity_N}) for different values of $\alpha^{2}$. Here, (a) $\alpha^{2}=4.0$ and (b) $\alpha^{2}=3.0$. Fidelity is used to quantify the closeness between the final state in equation (\ref{finalstate}) and the target state given by equation (\ref{target}).}
\label{fig:figure1}
\end{figure}

\begin{figure}[h]
\centering
\subfloat[]
{
    \includegraphics[scale=.5]{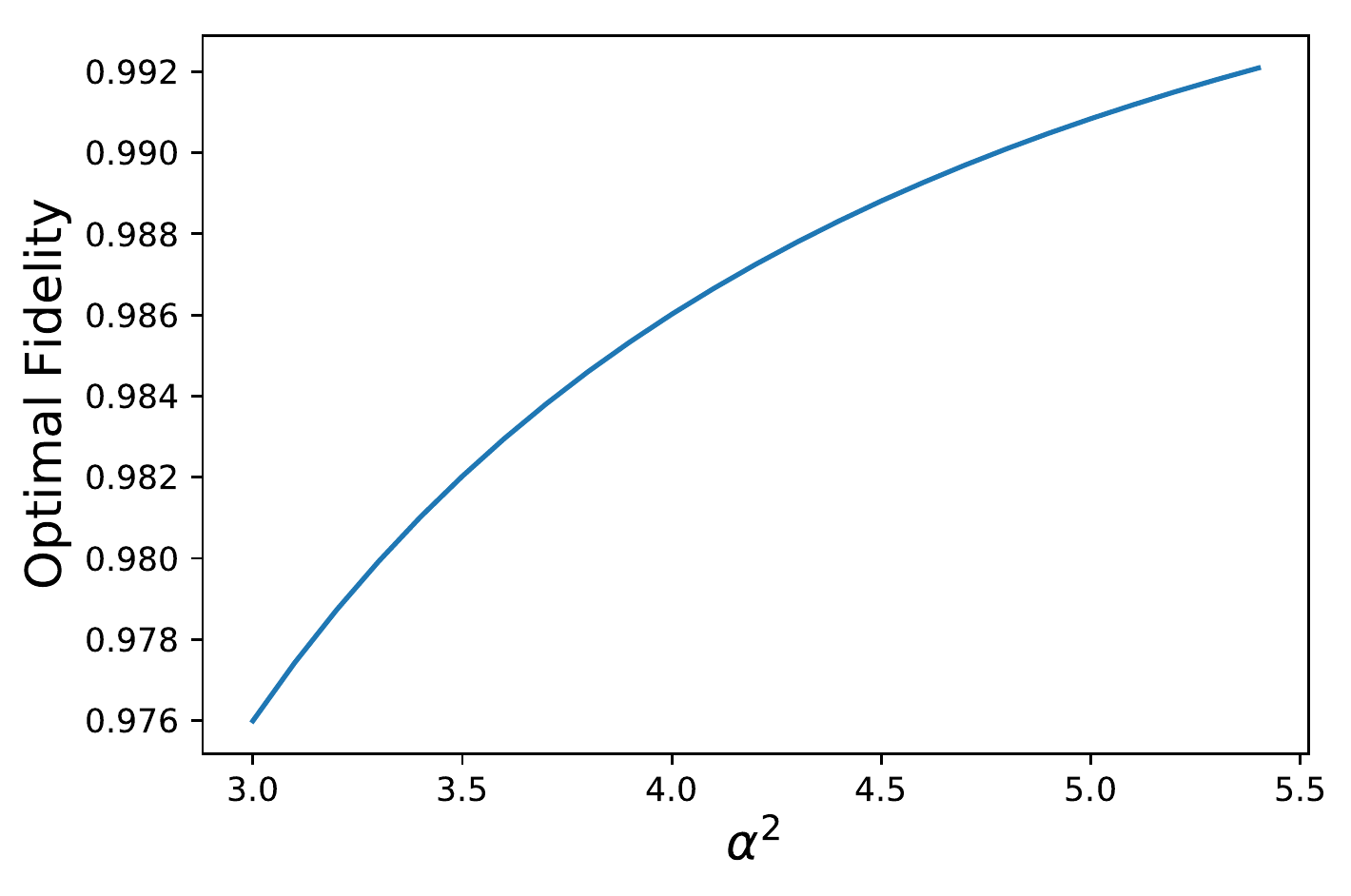}
    \label{fig:figure2a}
}
\subfloat[]
{
    \includegraphics[scale=.5]{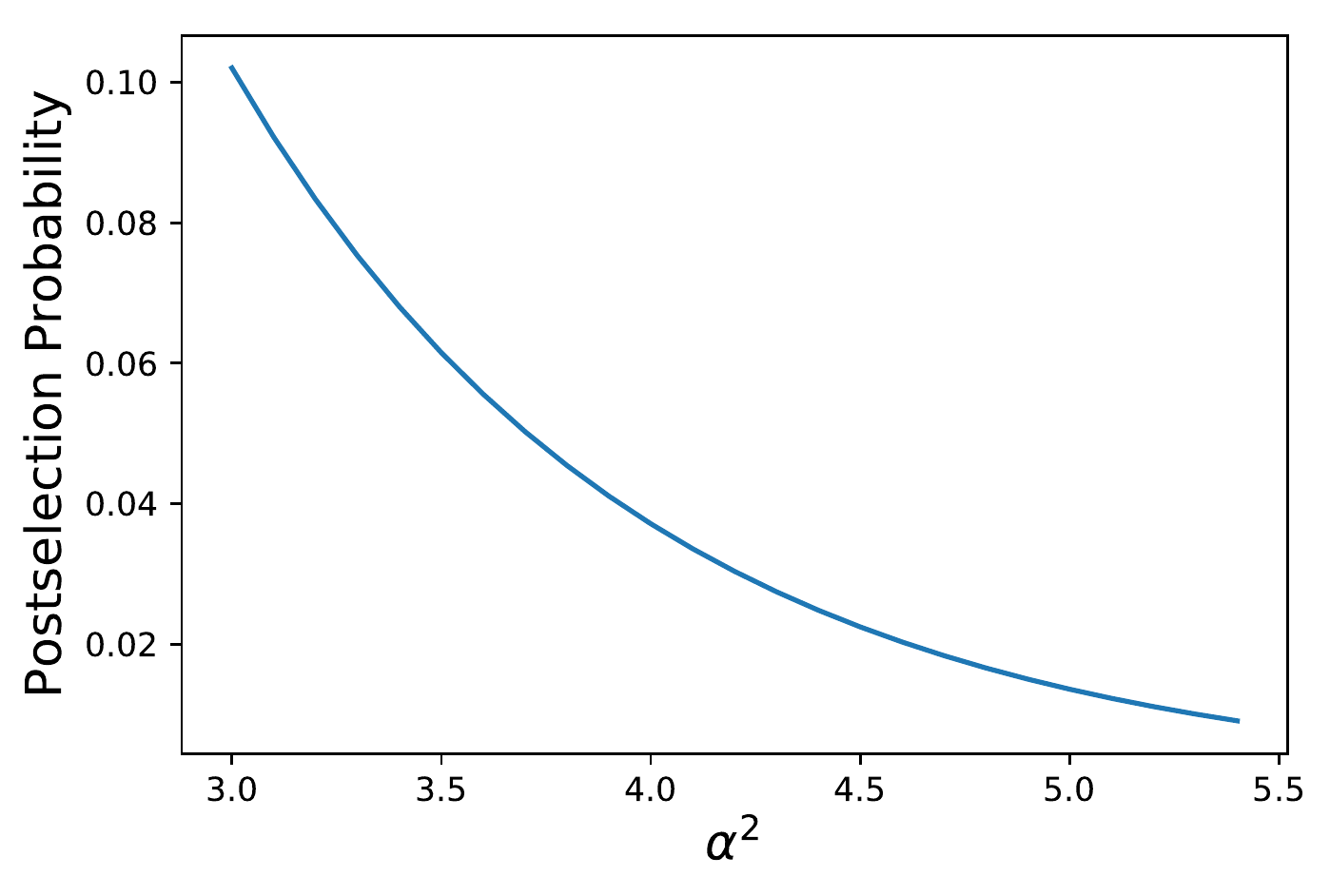}
    \label{fig:figure2b}
}
\caption{(a) Optimal fidelity from equation (\ref{fidelity_N}) versus $\alpha^{2}$. (b) Postselection probability for a set of parameters for which the fidelity is maximum (close to $1.0$).}
\label{fig:figure2}
\end{figure}

In the above equation, the number of atoms ($N$) has to be larger as $\alpha$ grows in order to maximize this fidelity. We plot the expression from equation (\ref{fidelity_N}) for different values of $\alpha^{2}$ in figure \ref{fig:figure1}. As we can see, for each $\alpha^{2}$ there is a number of atoms and a coupling value given by condition (\ref{phi}) for which the fidelity is optimal (close to $1.0$). In figure \ref{fig:figure2a} we determine the optimal fidelity $F_{opt}$ for each value of $\alpha^{2}$ and we also plot in figure \ref{fig:figure2b} the postselection probability given by equation (\ref{postprob}) for the optimal parameters. We found that for a range of 3.0 $<\alpha^{2}<$ 5.0, the optical qubit is generated with a fidelity and a postselection probability of $0.976 < F_{opt} <0.99$ and $10.2\% > P_{post} > 1.36\%$.

\begin{figure}[h]
\centering
\subfloat[]
{
    \includegraphics[scale=.4]{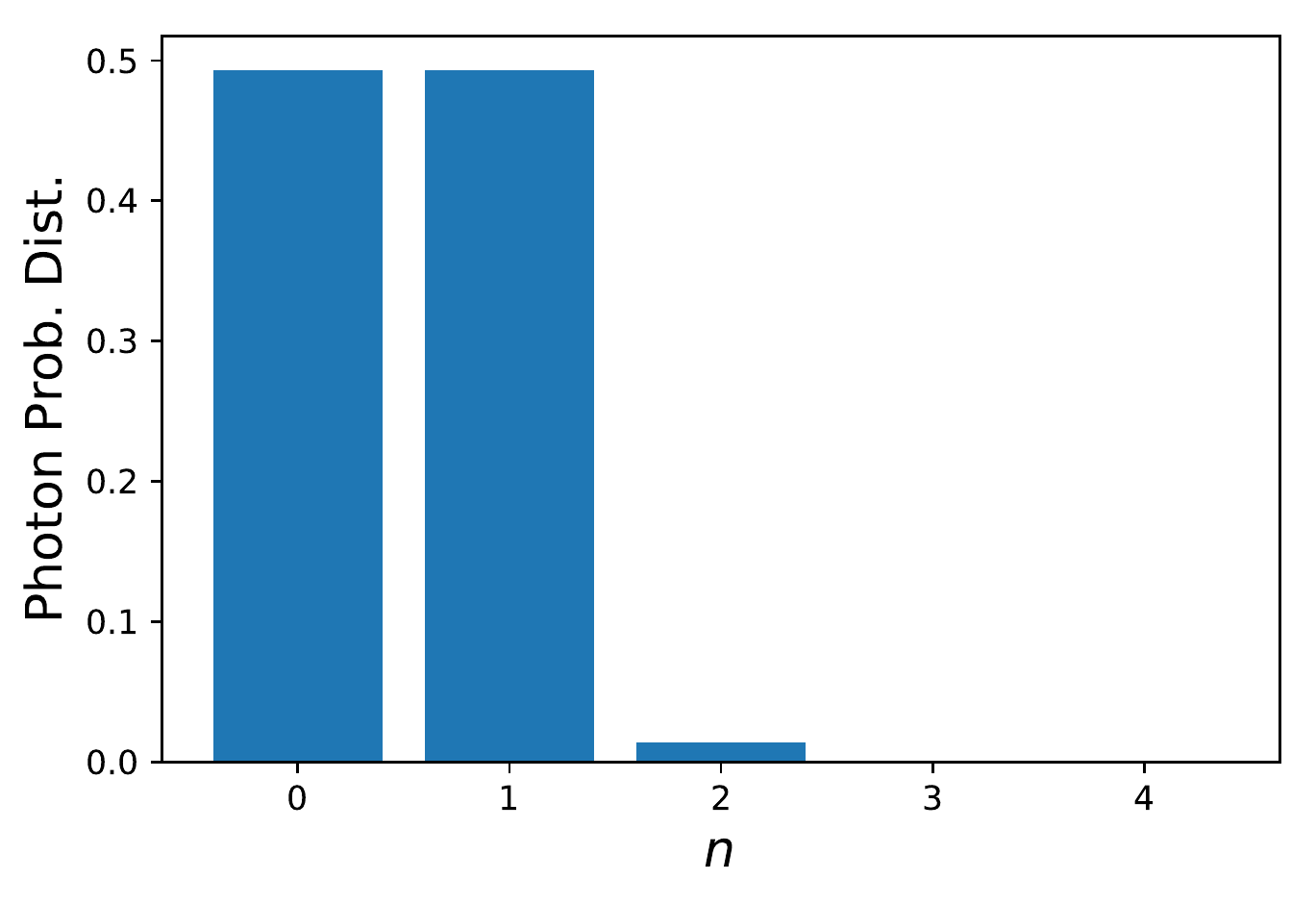}
    \label{fig:figure3a}
}
\subfloat[]
{
    \includegraphics[scale=.6]{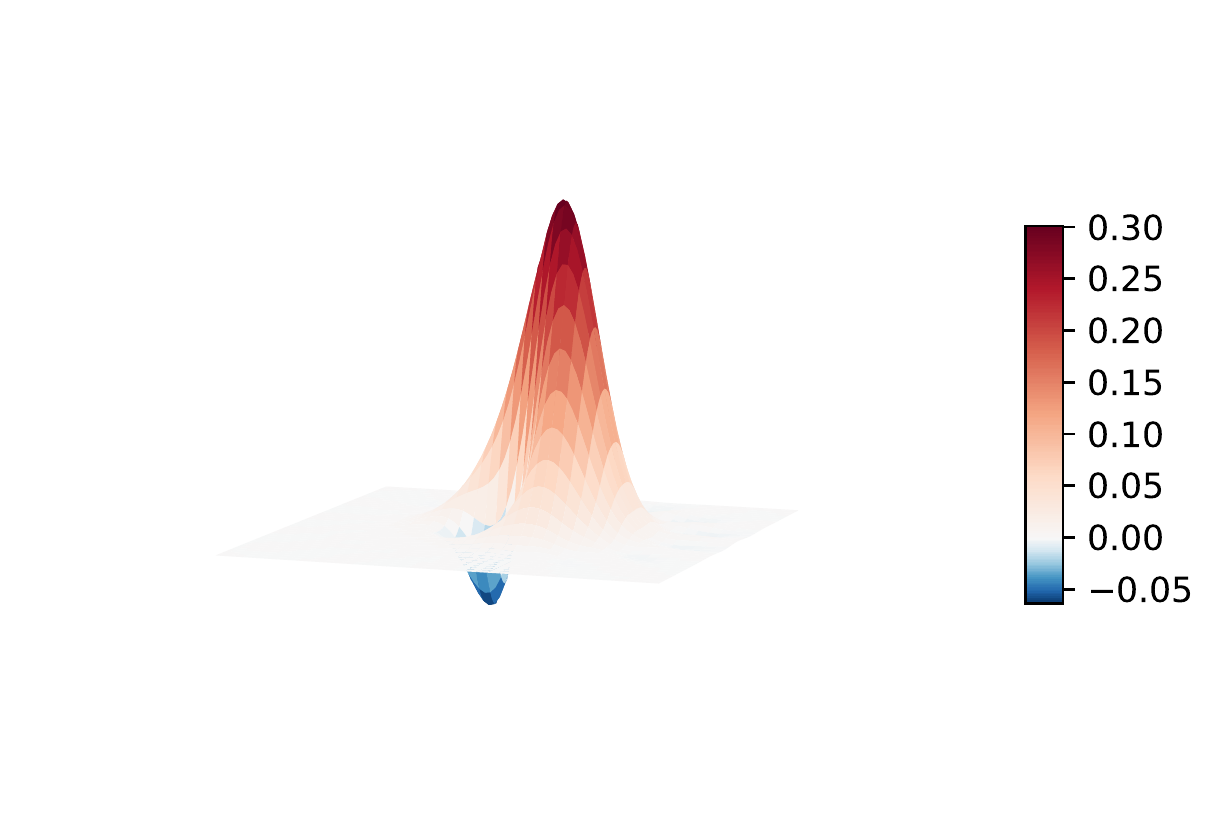}
    \label{fig:figure3b}
}
\caption{Generation of an optical qubit with $\alpha^{2}=4.0$, $N = 37$ and $\varphi\approx0.386$. (a) Photon probability distribution as in equation (\ref{photondist}). In (b) we show the Wigner function for the case described in (a). This optical qubit is generated with an optimal fidelity of $F_{opt}\approx0.986$ and a postselection probability of $P_{post}\approx3.72\%$.}
\label{fig:figure3}
\end{figure}

In figure \ref{fig:figure3a}, we show the probability distribution (Pr$(n)=|\langle n|\psi_{f}\rangle|^{2}$) of the final state given by equation (\ref{finalstate})
\begin{equation}\label{photondist}
Pr(n)=\frac{|b_{n}|^{2}\cos^{2N}\left(\frac{\varphi n}{2}\right)}{\sum_{n=0}^{\infty}|b_{n}|^{2}\cos^{2N}\left(\frac{\varphi n}{2}\right)}.
\end{equation}

We consider a set of parameters $\alpha$, $N$ and $\varphi$ that satisfy the condition (\ref{condition}) for an equiprobable superposition (Pr$(n=0)=$ Pr$(n=1)=1/2$) and for which the fidelity is maximum. However, the Hilbert space is not properly truncated up to just one photon, having Pr$(n=2)\approx1.38\%$ in the case shown in figure \ref{fig:figure3a}. To evidence the quantumness of the state, in figure \ref{fig:figure3b}, we have numerically computed the Wigner quasi-probability distribution defined as $W(x,p)=\frac{1}{\pi}\int_{-\infty}^{\infty}\langle x+x'|\psi_{f}\rangle\langle\psi_{f}|x-x'\rangle \e^{-2\rmi px'}\rmd x'$ \cite{wigner}. We observe that the true quantum nature arises as a consequence of the considerable negative part of $W(x,p)$.

To improve our results, we consider three groups of atoms crossing the cavities with three different couplings ($\varphi_{1}$, $\varphi_{2}$ and $\varphi_{3}$) between the atoms and the cavity field. All atoms are postselected in the ground state $|g\rangle$, giving a final state
\begin{equation}\label{finalstate2}
\fl|\psi_{f}\rangle = \frac{\sum_{n=0}^{\infty}b_{n}\e^{-\frac{\rmi n}{2}\left(\varphi_{1} N_{1}+\varphi_{2} N_{2}+\varphi_{3} N_{3}\right)}\cos^{N_{1}}\left(\frac{\varphi_{1} n}{2}\right)\cos^{N_{2}}\left(\frac{\varphi_{2} n}{2}\right)\cos^{N_{3}}\left(\frac{\varphi_{3} n}{2}\right)|n\rangle}{\left[\sum_{n=0}^{\infty}\left|b_{n}\right|^{2}\cos^{2N_{1}}\left(\frac{\varphi_{1} n}{2}\right)\cos^{2N_{2}}\left(\frac{\varphi_{2} n}{2}\right)\cos^{2N_{3}}\left(\frac{\varphi_{3} n}{2}\right)\right]^{1/2}}
\end{equation}
for the field after postselection, with a probability
\begin{equation}\label{postprob2}
\fl P_{post}=\sum_{n=0}^{\infty}\left|b_{n}\right|^{2}\cos^{2N_{1}}\left(\frac{\varphi_{1} n}{2}\right)\cos^{2N_{2}}\left(\frac{\varphi_{2} n}{2}\right)\cos^{2N_{3}}\left(\frac{\varphi_{3} n}{2}\right),
\end{equation}
where $N_{1}$, $N_{2}$ and $N_{3}$ are the number of atoms postselected with couplings $\varphi_{1}$, $\varphi_{2}$ and $\varphi_{3}$, respectively. Similarly to equation (\ref{target}), our target state is
\begin{equation}\label{target2}
\fl |\psi_{t}\rangle= \frac{|0\rangle+\alpha \e^{-\frac{\rmi}{2}\left(N_{1}\varphi_{1}+N_{2}\varphi_{2}+N_{3}\varphi_{3}\right)}\cos^{N_{1}}\left(\frac{\varphi_{1}}{2}\right)\cos^{N_{2}}\left(\frac{\varphi_{2}}{2}\right)\cos^{N_{3}}\left(\frac{\varphi_{3}}{2}\right)|1\rangle}{\sqrt{1+\alpha^{2}\cos^{2N_{1}}\left(\frac{\varphi_{1}}{2}\right)\cos^{2N_{2}}\left(\frac{\varphi_{2}}{2}\right)\cos^{2N_{3}}\left(\frac{\varphi_{3}}{2}\right)}}.
\end{equation}

Next, we consider $N_{1}=1$, $\varphi_{1}=\pi/2$, $N_{2}=1$ and $\varphi_{2}=\pi/3$ in order to kill the $n=2$ and $n=3$ components. As we can see from equation (\ref{finalstate2}) these values, $\varphi_{1}$ and $\varphi_{2}$, eliminated the kets $|2\rangle$ and $|3\rangle$, respectively. Using the same method applied before to prepare an optical qubit with an equiprobable superposition, we require
\begin{equation}\label{condition2}
\alpha \cos^{N_{3}}\left(\frac{\varphi_{3}}{2}\right) = 4/\sqrt{6}.
\end{equation}

Solving this condition for the variable $\varphi_{3}$, we have a fidelity depending on the number of atoms postselected in the $|g\rangle$ level given by
\begin{equation}\label{fidelity_N3}
\fl F(N_{3}) = \frac{2}{2+\sum_{n=2}^{\infty}\frac{\alpha^{2n}}{n!}\cos^{2}\left(\frac{\pi n}{4}\right)\cos^{2}\left(\frac{\pi n}{6}\right)\cos^{2N_{3}}\left(\frac{\varphi_{3} (N_{3})n}{2}\right)}.
\end{equation}

\begin{figure}[h] 
\centering
\subfloat[]
{
    \includegraphics[scale=.5]{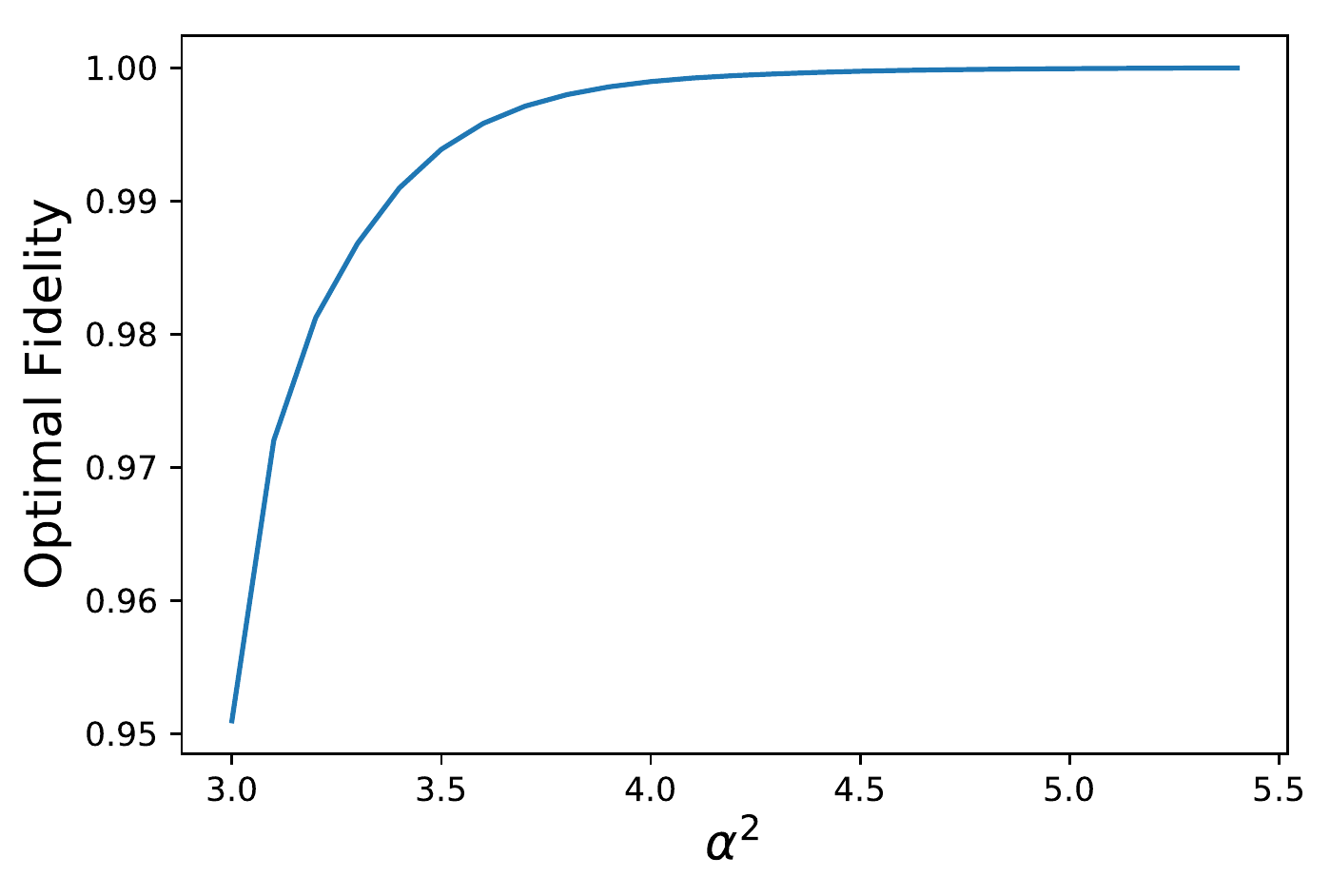}
    \label{fig:figure4a}
}
\subfloat[]
{
    \includegraphics[scale=.5]{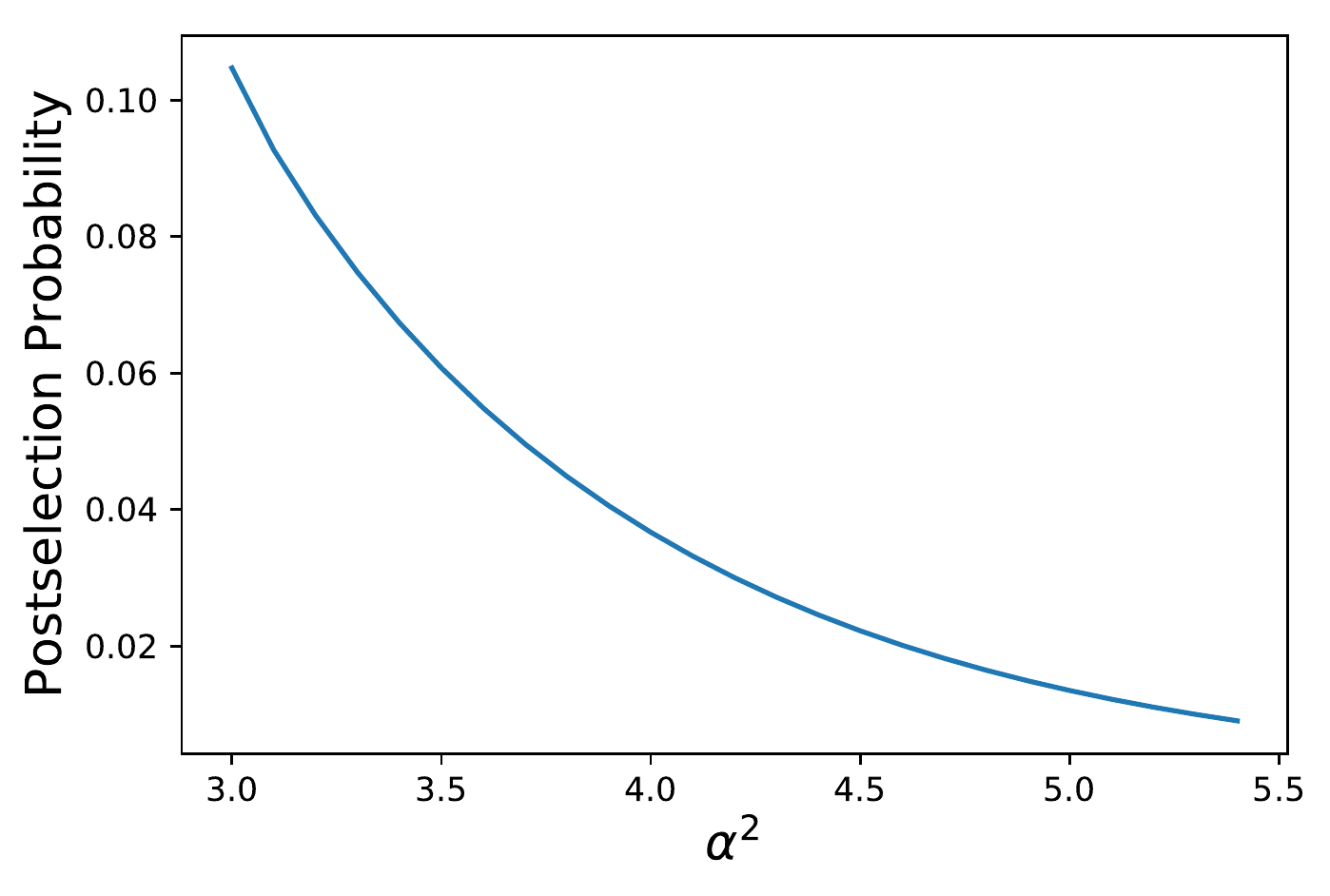}
    \label{fig:figure4b}
}
\caption{(a) Optimal fidelity from equation (\ref{fidelity_N3}) versus $\alpha^{2}$. (b) Postselection probability for a set of parameters $\alpha$, $N_{3}$ and $\varphi_{3}$ for which the fidelity is maximum.}
\label{fig: figure4}
\end{figure}

\begin{figure}[h]
\centering
\subfloat[]
{
    \includegraphics[scale=.4]{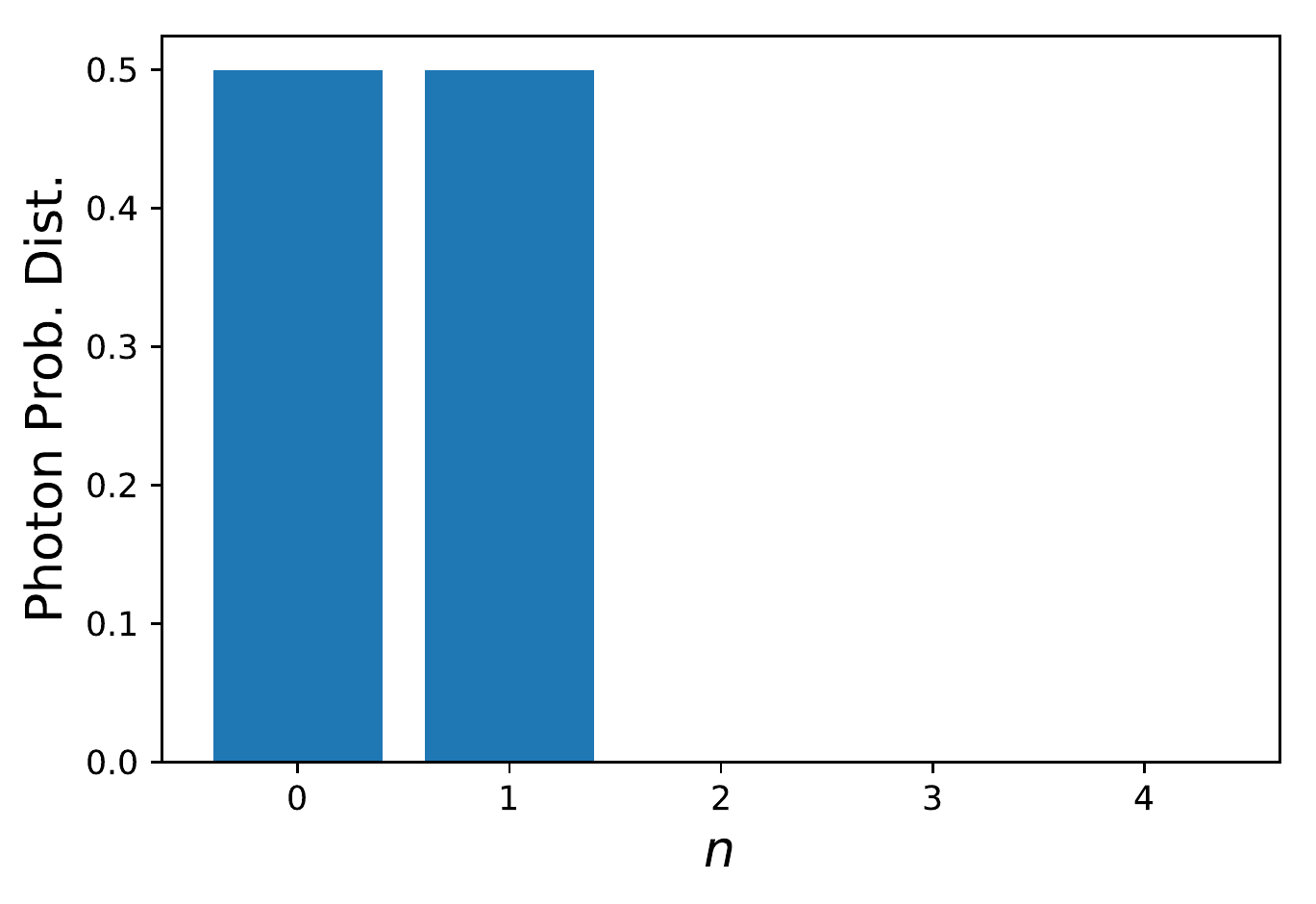}
    \label{fig:figure5a}
}
\subfloat[]
{
    \includegraphics[scale=.6]{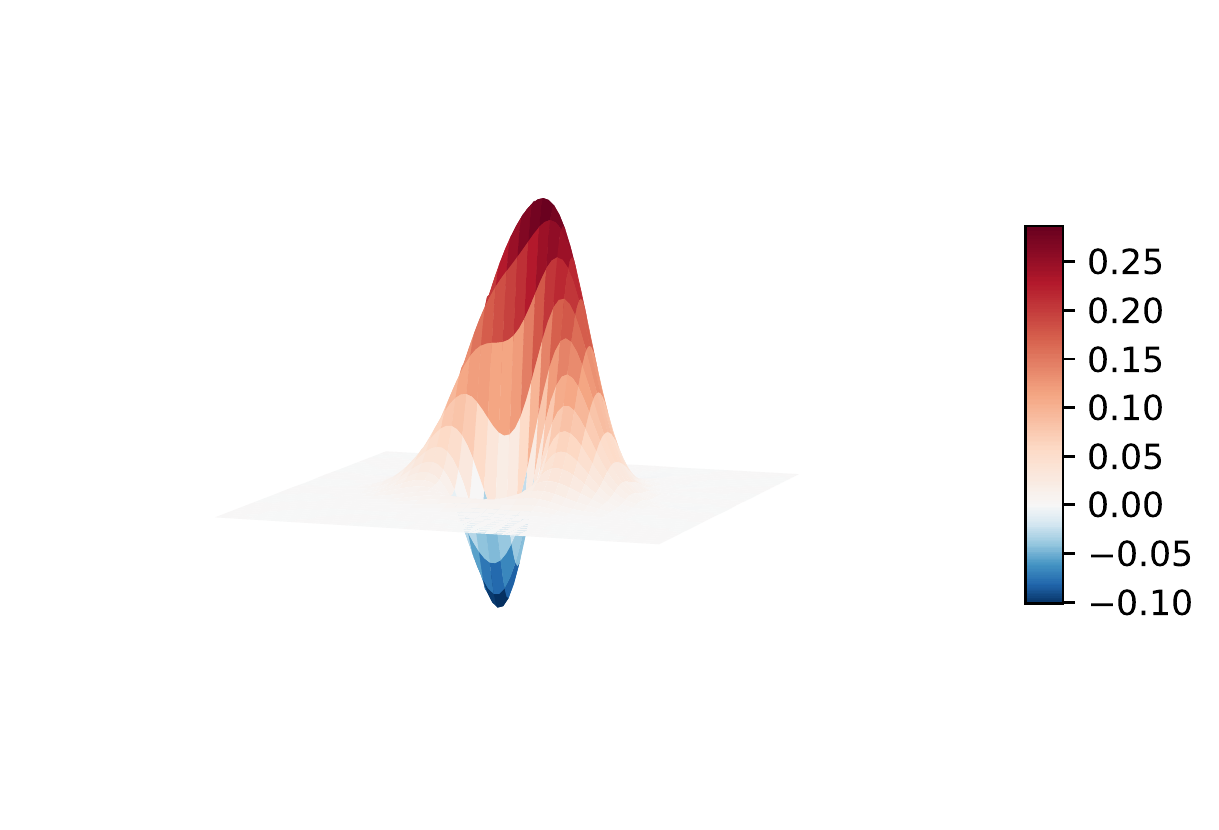}
    \label{fig:figure5b}
}
\caption{Generation of an optical qubit with $\alpha^{2}=4.0$, $N_{3}=11$ and $\varphi_{3}\approx0.383$. (a) Photon probability distribution of the postselected final state in equation (\ref{finalstate2}). In (b) we show the Wigner function for the case depicted in (a). This qubit is prepared with an optimal fidelity of $F_{opt}\approx0.999$ and a postselection probability of $P_{post}\approx3.67\%$.}
\label{fig:figure5}
\end{figure}

As before, the number of atoms ($N_3$) has to increase when $\alpha$ is larger in order to maximize the fidelity (equation (\ref{fidelity_N3})). Furthermore, the equation (\ref{fidelity_N3}) is maximized with less atoms than the previous case for a given value of $\alpha$. In figure \ref{fig:figure4a} we determine the optimal fidelity $F_{opt}$ for each value of $\alpha^{2}$ given that the other parameters satisfy condition (\ref{condition2}). Also, we plot in figure \ref{fig:figure4b} the postselection probability from equation (\ref{postprob2}) for the parameters for which the fidelity is maximum. In this case, we found that in the $3.0<\alpha^{2}<5.0$ range, the optical qubit is prepared with a fidelity and a postselection probability of $0.95<F_{opt}<0.999$ and $10.5\%>P_{post}>1.35\%$. Hence, we have improved the fidelity for the generation of the optical qubit keeping almost the same postselection probability.
We show in figure \ref{fig:figure5a} the photon probability distribution (Pr$(n)=|\langle n|\psi_{f}\rangle|^{2}$) calculated using the final state in equation (\ref{finalstate2}). We consider the same initial cavity field with $\alpha^2=4.0$ to show that we have improved the preparation of the equiprobable qubit having a very similar postselection probability to the previous case presented in figure \ref{fig:figure3a}. In figure \ref{fig:figure5b}, we have display the Wigner function to evidence the quantumness of the state presented in figure \ref{fig:figure5a}. It is also important to emphasize from the examples depicted in figure \ref{fig:figure3} and figure \ref{fig:figure5}, that we require less atoms to prepare the improved qubit, thus imply in a shorter interaction time and less dissipation.

The physical process that generates the superposition of the vacuum and one-photon states is the same involved in the reduction of the field into a Fock state in the QND procedure \cite{QND90, QND92, guerlin}. Essentially, after the continuous detection of the atoms, the field collapses into a coherent superposition of Fock states with amplitudes given by $b_{n}\e^{-\frac{\rmi n}{2}\left(\varphi_{1} N_{1}+\varphi_{2} N_{2}+\varphi_{3} N_{3}\right)}\cos^{N_{1}}\left(\frac{\varphi_{1} n}{2}\right)\cos^{N_{2}}\left(\frac{\varphi_{2} n}{2}\right)\cos^{N_{3}}\left(\frac{\varphi_{3} n}{2}\right)$ (within a normalization factor). Thus, the photon probability distribution is multiplied by an oscillating function of $n$. Consequently, the photon numbers for which this function is close to zero are efficiently decimated. However, in our work the decimation process is not random because we determine the parameters that efficiently decimate all the photon numbers except $n=0$ and $n=1$. Therefore, the most important ingredient is postselection. 
\section{Discussion and conclusion}
\subsection{A more realistic scenario}
The analysis described above is most valid mostly in an idealized experiment. In the present section we describe a more realistic scenario and include the effects of experimental constraints such as cavity losses and imperfect detection of the atoms due to detection efficiency and error detection. Additionally, we present some examples of the generation of superpositions of higher-photon-number Fock states using this scheme. 

This experiment can be done in a typical microwave QED system. Here the atoms sent across the cavity are circular Rydberg atoms. This kind of atoms have a very long lifetime, on the order of tens of miliseconds, comparable to the lifetime of the photon in the superconducting cavity (130 miliseconds) with a fully open structure needed for passing the atoms through. So we can neglect the atomic decay process during the interaction time between the atoms and the cavity field, and also consider that the atoms fly coherently through the cavities due to the short interaction time ($\sim$ 0.4 miliseconds). All of the parameters of the atomic samples are controllable (velocity, preparation time, interaction time, etc). Therefore, the different couplings $\varphi_{k}$ needed in our scheme to prepare the qubit with high fidelity can be realized by controlling individually the velocity of the atoms by laser techniques. Previously, we mentioned that we need only one atom in the setup at a given time. However, in real experiments, it is not readily possible to handle a deterministic single-atom preparation. One way to emulate single-atom experiments is preparing the atoms by weak laser excitation, producing a Poissonian statistics for the atoms (with a mean number of atoms per sample much less than one). Then, a postselection process takes place in which we retain only the data corresponding to the desired state. The single-atoms events are obtained with an increase in the time of the data acquisition \cite{exploring}. Nevertheless, we assume in our study a deterministic single atom preparation and there are some proposals to achieve this preparation of Rydberg atoms making use of the called \textit{dipole blockade} effect \cite{dipoleblockade1, dipoleblockade2}. 

In a typical experiment, the field is stored in a superconducting cavity $C$ (cavity damping time $T_{c}=65$ ms) cooled down to a temperature $T=0.8$ K \cite{kuhr}, and its dynamics is described by the master equation for a reservoir at temperature $T$ \cite{scully}
\begin{eqnarray}
\nonumber\frac{\rmd\rho}{\rmd t}=\textbf{L}\rho=-\frac{\kappa}{2}(1+&n_{th})(a^{\dag}a\rho+\rho a a^{\dag}-2a\rho a^{\dag})
\\
&-\frac{\kappa}{2}n_{th}(a a^{\dag}\rho+\rho a^{\dag}a-2a^{\dag}\rho a),
\label{masterequation}
\end{eqnarray}
where $\kappa=1/T_{cav}$ is the cavity decay rate and $n_{th}=0.05$ is the equilibrium thermal photon number. The atoms are sent at a $T_a=82$ $\mu$s time interval \cite{sayrin}. Within the approximation of small time interval, $T_a\ll T_c$, we can describe the evolution of the field due to the cavity field relaxation during the time interval $T_a$ between two atoms by the action of the superoperator $\textbf{T}$ \cite{dotsenko}:
\begin{equation}
\textbf{T}\rho=(\mathbb{I}+T_{a}\textbf{L})\rho.
\label{T}
\end{equation}

As we mentioned before the atomic detection is not perfect. The detector $D$ has a finite detection efficiency $\eta_{d}$ (probability of detecting an atom). Additionally, the limited state resolution of the Ramsey interferometer introduces a detection error probability of $\eta_{f}$. In our calculations we use $\eta_{d}=0.87$ and $\eta_{f}=0.05$.

Because of nonideal detection efficiency and nonzero effective detection errors, a measurement outcome $m'=e$ or $g$ corresponds to a statistical mixture of different ideal measurement outcomes $m$. The conditional probabilities $P(m'|m)$ and the ideal detection operators $M_{m'}$ are given in \cite{peaudecerf}. We now give the explicit expression of a superoperator $\textbf{P}_{m'}$ acting on $\rho$ describing the imperfect detection of an atom 
\begin{equation}
\textbf{P}_{m'}\rho=\frac{\sum_{m}P(m'|m)M_{m} \rho M_{m}^{\dag}}{\Tr\left(\sum_{m}P(m'|m)M_{m} \rho M_{m}^{\dag}\right)}.
\label{P}
\end{equation}

\begin{figure}[h]
\centering
\subfloat[]
{
    \includegraphics[scale=.5]{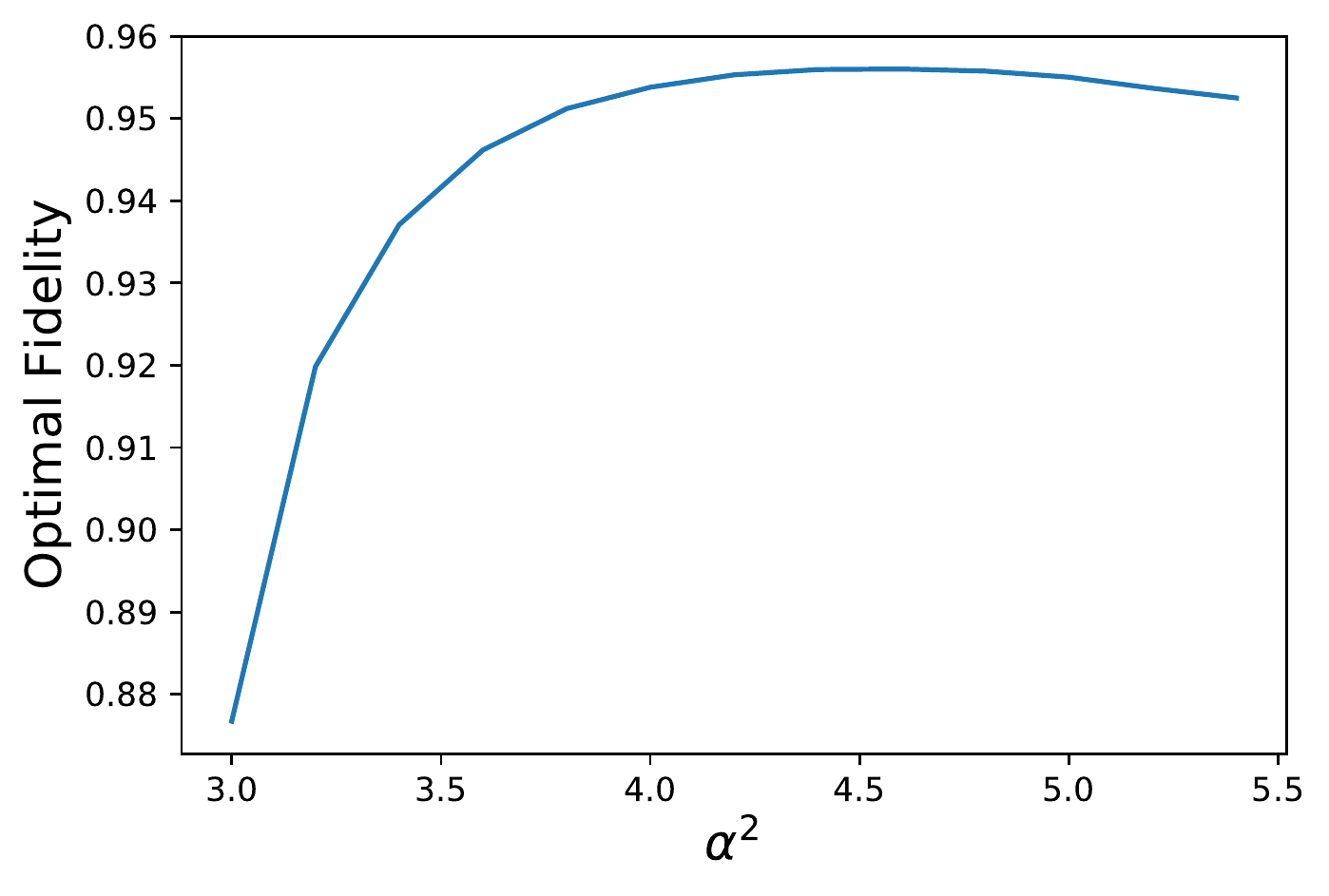}
    \label{fig:figure6a}
}
\subfloat[]
{
    \includegraphics[scale=.5]{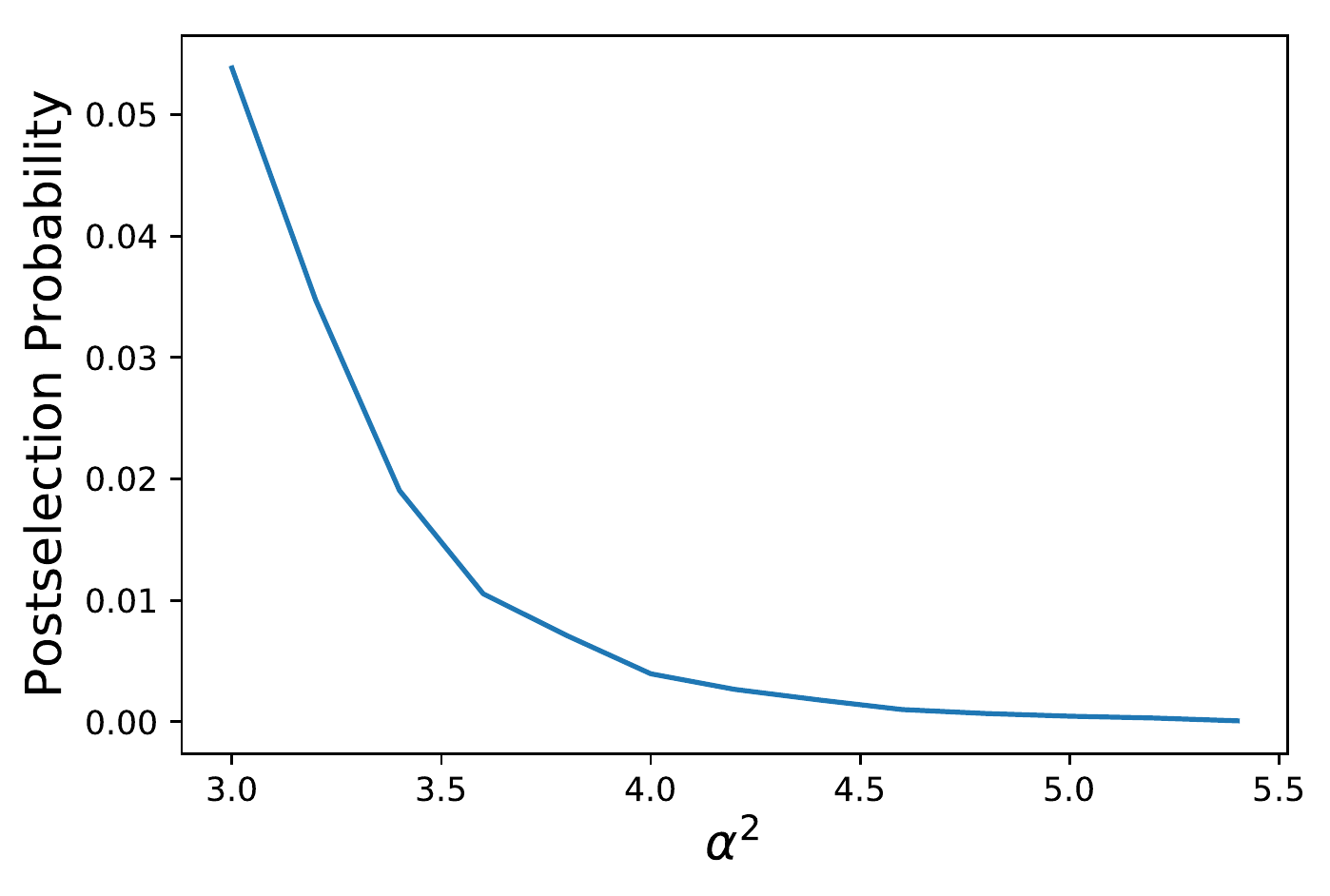}
    \label{fig:figure6b}
}
\caption{(a) Optimal fidelity including the effect of the cavity relaxation for each $\alpha^{2}$. (b) Post-selection probability for a set of parameters $\alpha$, $N_{3}$ and $\varphi_{3}$ for which the fidelity is maximum.}
\label{fig: figure6}
\end{figure}

In our study to generate the optical qubit select the measurement outcome $m'=g$, using equation (\ref{P}) the detection of this outcome is
\begin{equation}
\textbf{P}_{g}\rho=\frac{\eta_{d}(1-\eta_{f})M_{g}\rho M_{g}^{\dag}+\eta_{d}\eta_{f}M_{e}\rho M_{e}^{\dag}}{\eta_{d}(1-\eta_{f})\Tr(M_{g}\rho M_{g}^{\dag})+\eta_{d}\eta_{f}\Tr(M_{e}\rho M_{e}^{\dag})}
\label{detection}
\end{equation}
As before, our initial state is coherent and the target is the pure state given by equation (\ref{target2}). Requiring that all the atoms are detected in $|g\rangle$, we include the effect of the cavity relaxation between each detection using equation (\ref{T}) and (\ref{detection}). We optimize the fidelity defined as $F = \langle\psi_{t}|\rho_{f}|\psi_{t}\rangle$ \cite{fidelity}, where the final state after the postselection is represented by the density matrix $\rho_f$ due to the effect of the cavity losses and imperfect detections. In figure \ref{fig:figure6a} we show the optimal fidelity versus $\alpha^{2}$ with the parameters $\varphi_{3}$ and $N_{3}$ satisfying condition (\ref{condition2}). Also, in figure \ref{fig:figure6b} the postselection probability is shown. In figure \ref{fig:figure6a} we observe that as $\alpha^2$ increases, the fidelity becomes smaller as compared to the result without photon losses (figure \ref{fig:figure4a}). The reason behind this is that larger $\alpha^2$ is also translated into more atomic postselection steps ($N$) required to generate the optical qubit, i.e. longer interaction times are needed for our scheme to work. Therefore, the effects of photons leaking from the cavity are more probable for larger $\alpha^2$. As we can see from figure \ref{fig:figure6b} the postselection probability is significantly reduced. However, we have a preparation of the optical qubit with a fidelity and postselection probability $0.9<F_{opt}<0.94$ and $4.28\%>P_{post}>1.72\%$, which is still within experimental reach.

At this stage, we have considered a more realistic scenario including the effect of the cavity losses, detection efficiency ($\eta_{d}<1$) and error detection ($\eta_{f}>0$), obtaining a robust preparation of the optical qubit in a real Fabry-P\'{e}rot superconducting cavity. 
\subsection{Other superpositions of photon number Fock states}
\begin{figure}[h]
\centering
\subfloat[]
{
    \includegraphics[scale=.4]{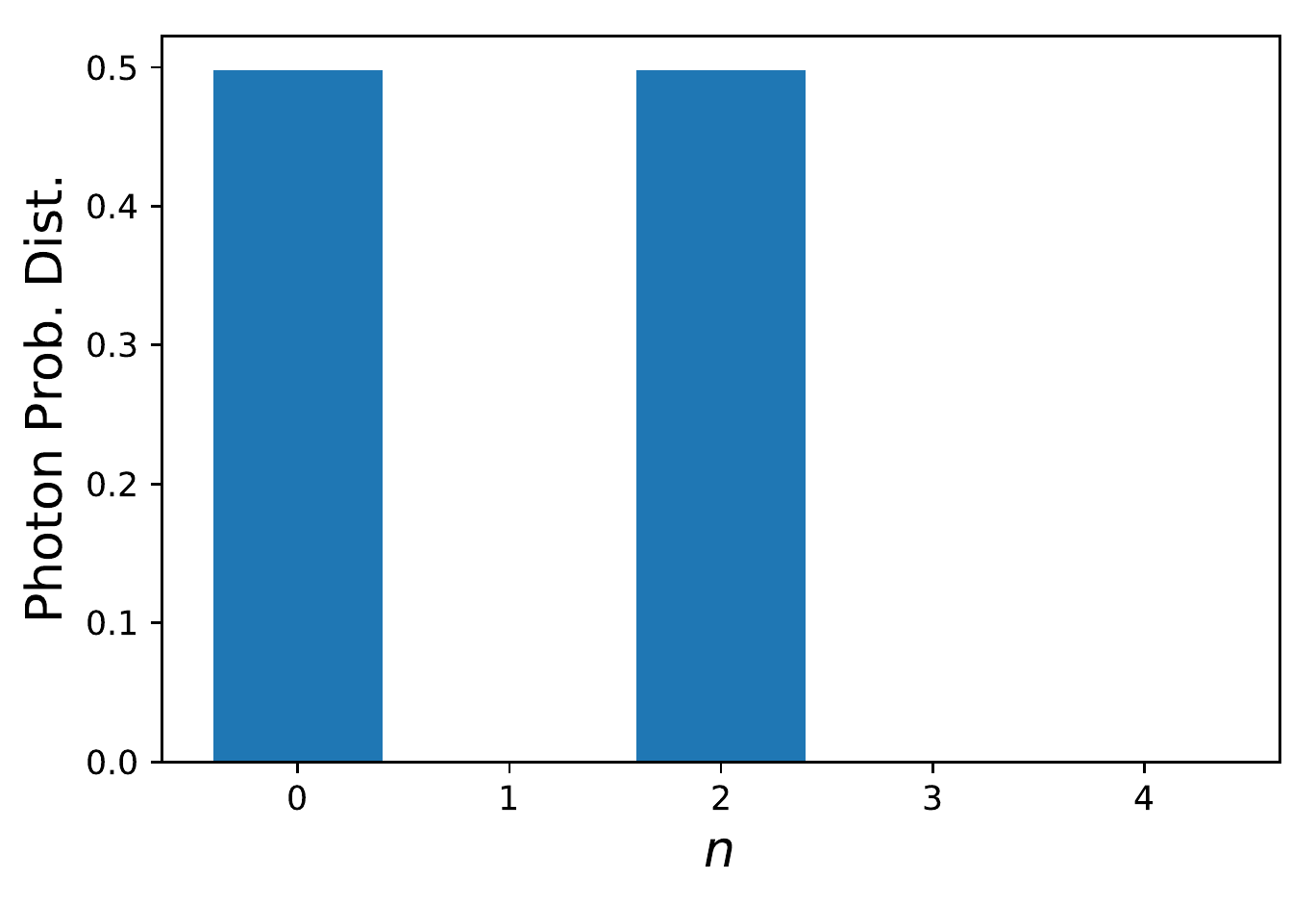}
    \label{fig:figure8a}
}
\subfloat[]
{
    \includegraphics[scale=.6]{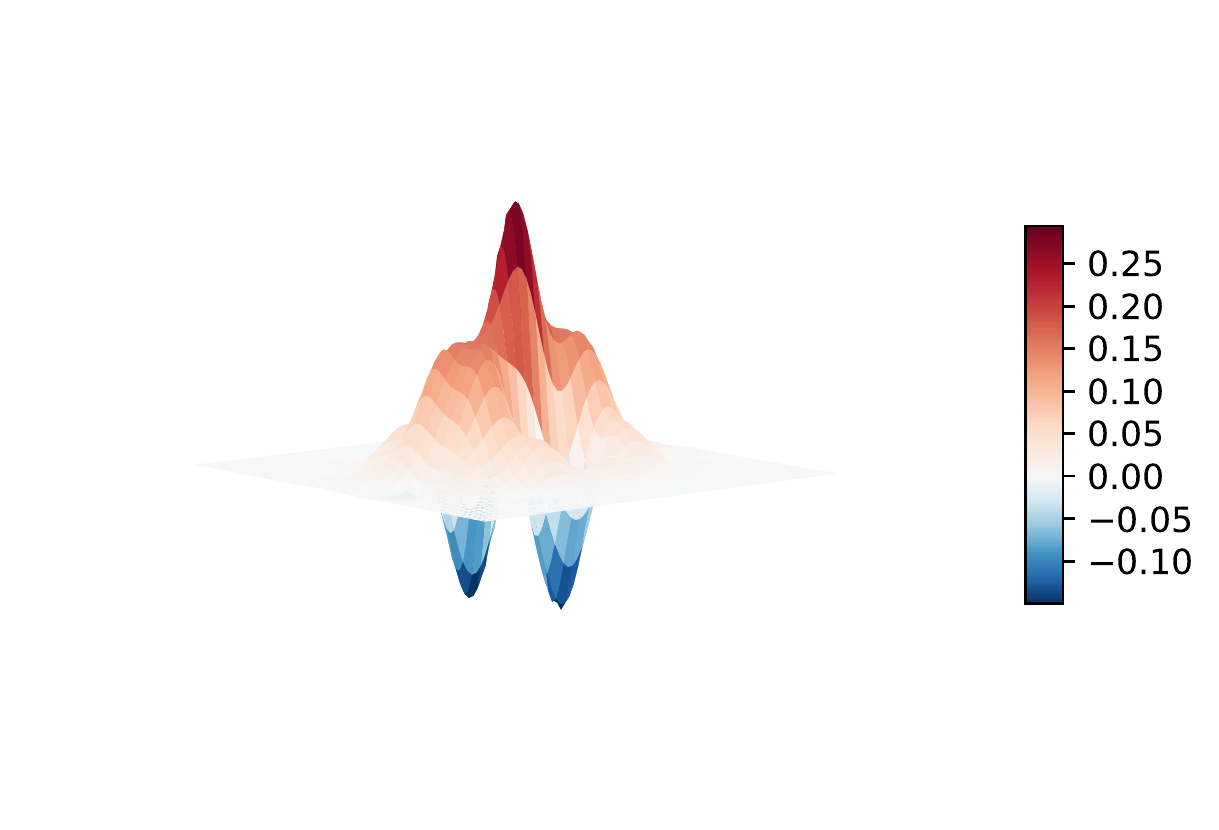}
    \label{fig:figure8b}
}

\subfloat[]
{
    \includegraphics[scale=.4]{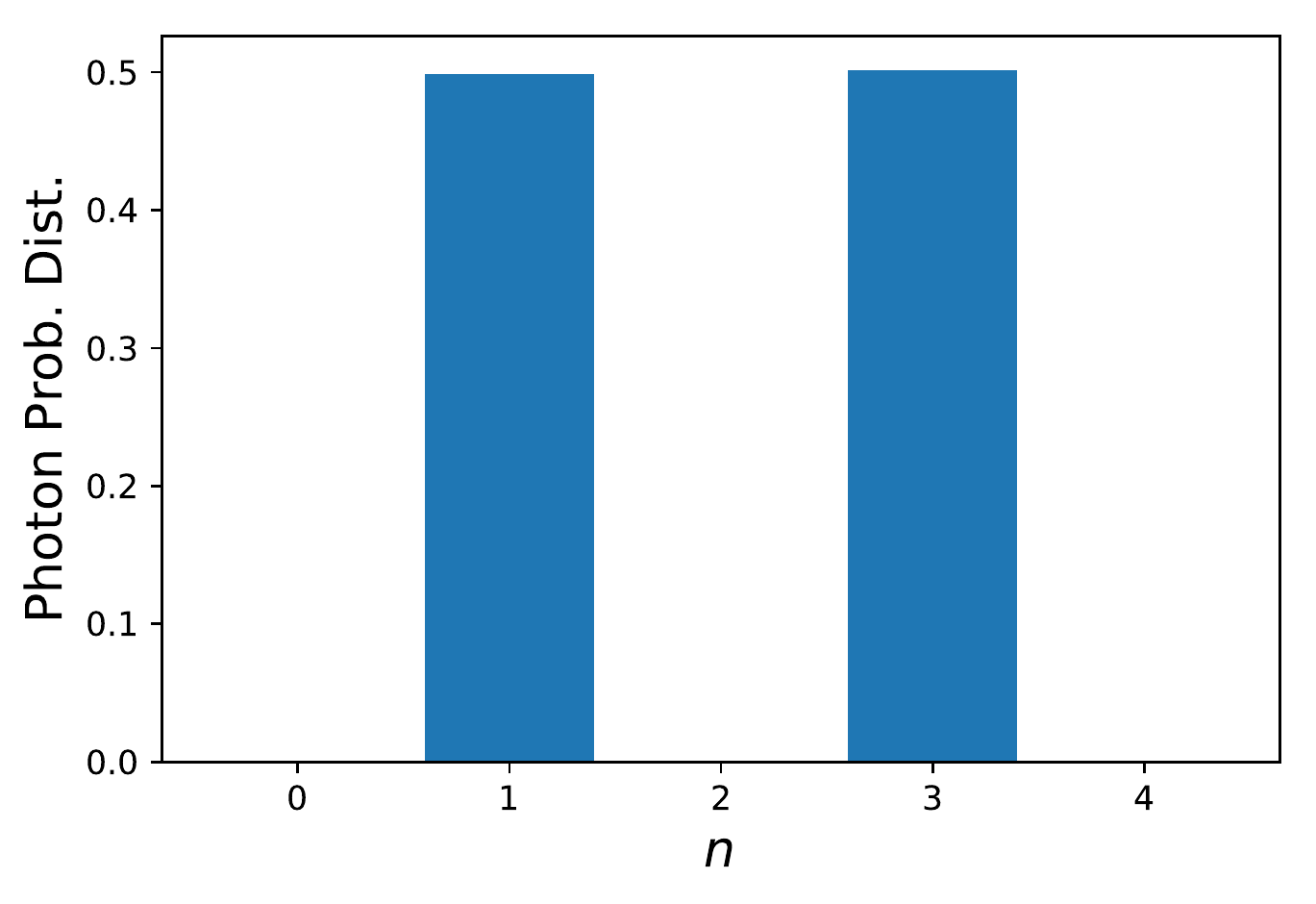}
    \label{fig:figure9a}
}
\subfloat[]
{
    \includegraphics[scale=.6]{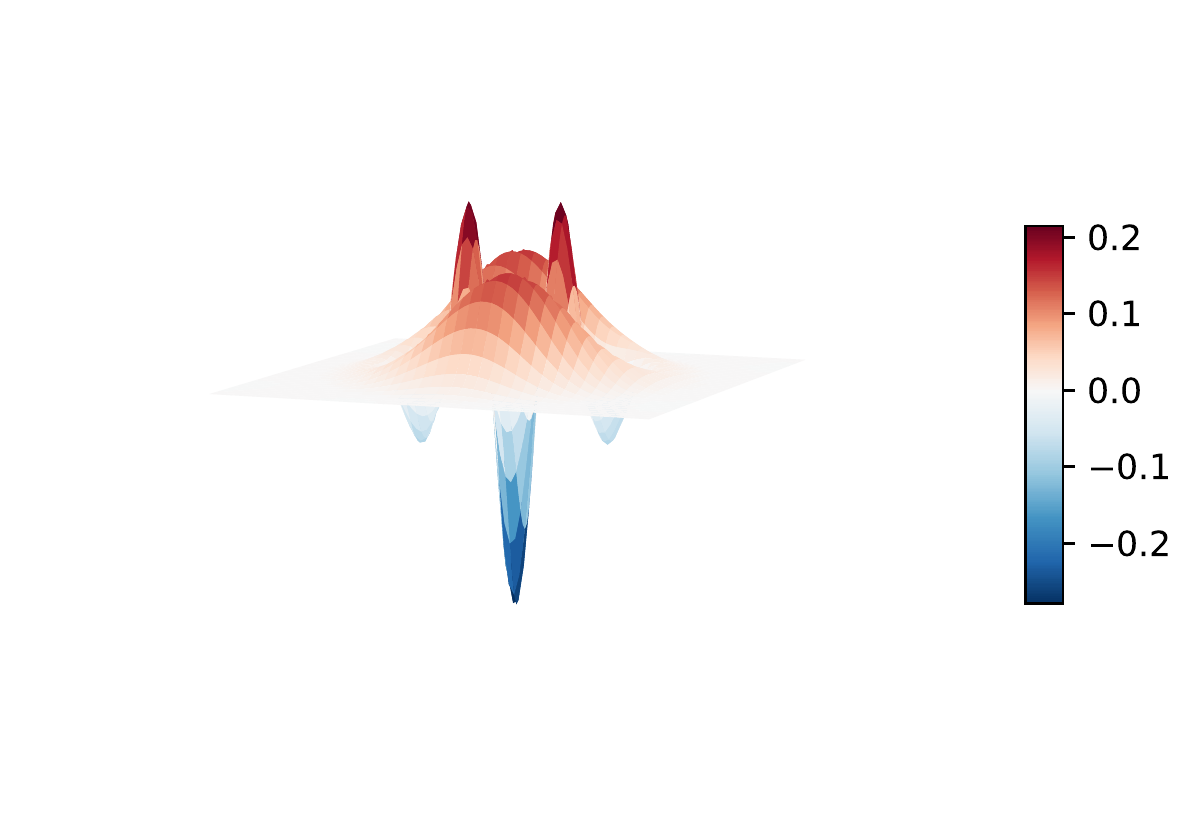}
    \label{fig:figure9b}
}
\caption{Other superpositions of photon number Fock states. In (a) and (b) we show a superposition of states $|0\rangle$ and $|2\rangle$. This state is prepared with a fidelity of 0.99 and a probability of $10\%$. Figures (b) and (c) show a superposition of states $|1\rangle$ and $|3\rangle$. This state is prepared with a fidelity of 0.97 and a probability of $5.5\%$.}
\label{fig:figure8}
\end{figure}

Finally, if we postselect not only atoms in the $|g\rangle$ level it is possible to generate other superpositions of photon number Fock states using this scheme. A more general expression for a pure state is derived from equation (\ref{finalstate}), giving us
\begin{equation}
|\psi_{f}\rangle = \frac{\sum_{n=0}^{\infty}b_{n}\e^{-\frac{\rmi n\varphi N}{2}}\cos^{N-N_{e}}\left(\frac{\varphi n}{2}\right)\sin^{N_{e}}\left(\frac{\varphi n}{2}\right)|n\rangle}{\left[\sum_{n=0}^{\infty}\left|b_{n}\right|^{2}\cos^{2(N-N_{e})}\left(\frac{\varphi n}{2}\right)\sin^{2N_{e}}\left(\frac{\varphi n}{2}\right)\right]^{1/2}}.
\end{equation}
Therefore, adjusting the number of atoms detected in each level ($N_{e}$ in $|e\rangle$ and  $N-N_{e}$ in $|g\rangle$) and their interactions ($\varphi$), we determine the parameters that decimate other photon number states to properly generate higher photon number Fock state superpositions. In figure \ref{fig:figure8a} and \ref{fig:figure8b}, we show a superposition of $|0\rangle$ and $|2\rangle$ states. First, one atom interacting with $\varphi_{1}=\pi$ is detected in $|g\rangle$, then five atoms interacting with $\varphi_{2} = 0.535$ are detected in $|g\rangle$. Also, we include the superposition of $|1\rangle$ and $|3\rangle$ states in figure \ref{fig:figure9a} and \ref{fig:figure9b}. First, two atoms interacting with $\varphi_{1}=\pi$ and $\varphi_{3}=\pi/5$  are detected in $|e\rangle$ and $|g\rangle$ , respectively. Then, with $\varphi_{3}=0.372$ we detected one atom in $|e\rangle$ and four atoms in $|g\rangle$.

In summary, we suggest for the first time a scheme to generate an optical qubit from an initial coherent state of the field in a typical cavity QED setup using a dispersive atom-field interaction and postselection of atoms. Particularly, we study the case of an equiprobable superposition of the vacuum and one-photon states. First, the general scheme for the generation of a cavity field state from an initial state via atomic postselection is presented. Then, we focus on the preparation of the optical qubit by setting the parameters which optimize the fidelity between the final and our desired state. As seen from the previous sections, we can achieve this goal with a high fidelity and a postselection probability within experimental reach. Finally, we conclude our study showing that this scheme can generate other superpositions of photon number Fock states.
\bigskip
\section*{Acknowledgments}
M.O. acknowledges the financial support of the project Fondecyt $\#$1180175, and F.O. acknowledges the financial support of the program Conicyt Beca Mag\'{i}ster Nacional 22172133.


\section*{References}
\begin{thebibliography}{}
\bibitem{krause} Krause J, Scully M O, Walther T and Walther H 1989 {\it Phys. Rev.} A {\bf 39} 1915

\bibitem{QND90} Brune M, Haroche S, Lefevre V, Raimond J M and Zagury N 1990 {\it Phys. Rev. Letter.} {\bf 65} 976

\bibitem{QND91} Holland M J, Walls D F and Zoller P 1991 {\it Phys. Rev. Letter.} {\bf 67} 1716

\bibitem{QND92} Brune M, Haroche S, Lefevre V, Raimond J M, Davidovich L and Zagury N 1992 {\it Phys. Rev.} A {\bf 45} 5193

\bibitem{leonski} Leo\'{n}ski W 1996 {\it Phys. Rev.} A {\bf 54} 3369

\bibitem{leonski2} Leo\'{n}ski W 1997 {\it J. Mod. Opt.} {\bf 44} 2105
 
\bibitem{domokos} Domokos P, Brune M, Raimond J M and Haroche S 1998 {\it Eur. Phys. J.} D {\bf 1} 1


\bibitem{varcoe} Varcoe B T H, Brattke S, Weidinger M and Walther H 2000 {\it Nature} {\bf 403} 743

\bibitem{dotsenko} Dotsenko I, Mirrahimi M, Brune M, Haroche S, Raimond J M and Rouchon P 2009 {\it Phys. Rev.} A {\bf 80} 013805 

\bibitem{voguel} Voguel K, Akulin V M and Schleich W P 1993 {\it Phys. Rev. Letter.} {\bf 71} 1816

\bibitem{moussa} Moussa M H Y and Baseia B 1998 {\it Phys. Lett.} A {\bf 238} 223

\bibitem{PPB} Pegg D T, Phillips L S and Barnett S M 1998 {\it Phys. Rev. Letter.} {\bf 81} 1604 

\bibitem{Barnett} Barnett S M and Pegg D T 1999 {\it Phys. Rev.} A {\bf 60} 4965

\bibitem{dakna} Dakna M, Clausen J, Knoll L and Welsch D G 1999 {\it Phys. Rev.} A {\bf 59} 1658

\bibitem{paris} Paris M G A 2000 {\it Phys. Rev.} A {\bf 62} 033813

\bibitem{dariano} D'Ariano G M, Maccone L, Paris M G A and Sacchi M F 2000 {\it Phys. Rev.} A {\bf 61} 053817

\bibitem{serra} Serra R M, de Almeida N G, Villas-B\^{o}as C J and Moussa M H Y 2000 {\it Phys. Rev.} A {\bf 62} 043810

\bibitem{qsd2001} \"{O}zdemir {S} K, Miranowicz A, Koashi M and Imoto N 2001 {\it Phys. Rev.} A {\bf 64} 063818

\bibitem{qsd2002} \"{O}zdemir {S} K, Miranowicz A, Koashi M and Imoto N 2002 {\it J. Mod. Opt.} {\bf 49} 977

\bibitem{nonlinearscissors} Miranowicz A, Ozdemir S K,  Leonski W, Koashi M and Imoto N 2003 {\it 13th Polish-Czech-Slovak Conference on Wave and Quantum Aspects of Contemporary Optics} vol 5259 (Bellingham: SPIE)

\bibitem{guerlin} Guerlin C {\it et al} 2007 {\it Nature} {\bf 448} 889

\bibitem{sayrin} Sayrin C {\it et al} 2011 {\it Nature} {\bf 477} 73

\bibitem{resch} Resch K J, Lundeen J S and Steinberg A M 2002 {\it Phys. Rev. Letter.} {\bf 88} 113601

\bibitem{scully} Scully M O and Zubairy M S 1997 {\it Quantum Optics} (New York: Cambridge University Press)

\bibitem{fidelity} Jozsa R 1994 {\it J. Mod. Opt.} {\bf 41} 2315

\bibitem{wigner} Leonhardt U 1997 {\it Measuring the Quantum State of Light} (New York: Cambridge University Press)

\bibitem{exploring} Haroche S and Raimonf J M 2006 {\it Exploring the Quantum: Atoms, Cavities and Photons} (New York: Oxford University Press)

\bibitem{dipoleblockade1} Lukin M D {\it et al} 2001 {\it Phys. Rev. Letter.} {\bf 87} 037901

\bibitem{dipoleblockade2} Saffman M and Walker T G 2002 {\it Phys. Rev.} A {\bf 66} 065403


\bibitem{kuhr} Kuhr S {\it et al} 2007 {\it Appl. Phys. Letters.} {\bf 90} 164101

\bibitem{peaudecerf} Peaudecerf B {\it et al} 2013 {\it Phys. Rev.} A {\bf 87} 042320
\end{thebibliography}
\end{document}